\documentclass[lettersize,journal]{IEEEtran}
\usepackage{amsmath,amsfonts}
\usepackage{algorithmic}
\usepackage{algorithm}
\usepackage{array}
\usepackage[caption=false,font=normalsize,labelfont=sf,textfont=sf]{subfig}
\usepackage{textcomp}
\usepackage{stfloats}
\usepackage{url}
\usepackage{verbatim}
\usepackage{graphicx}
\usepackage{cite}
\usepackage{xcolor}
\usepackage{booktabs}
\usepackage{threeparttable}
\usepackage{colortbl}
\usepackage{multirow}

\usepackage{makecell}
\usepackage{pifont}

\definecolor{dark-green}{HTML}{228B22}
\definecolor{dark-blue}{HTML}{1976D2}
\definecolor{dark-purple}{HTML}{8d4bbb}
\definecolor{dark-red}{HTML}{D63C3C}
\usepackage{hyperref}
\hypersetup{
    colorlinks=true,
    linkcolor=dark-red,
    filecolor=magenta,      
    urlcolor=dark-red,
    citecolor=dark-blue,
}

\newcommand{\shh}[1]{\textcolor{black}{#1}}

\newcommand{\lxy}[1]{\textcolor{black}{#1}}
\newcommand{\lxyRR}[1]{\textcolor{black}{#1}}
\newcommand{\lxyMinor}[1]{\textcolor{black}{#1}}
\newcommand{\lxyMajor}[1]{\textcolor{black}{#1}}

\newcommand{\cmark}{\ding{51}}%
\newcommand{\xmark}{\ding{55}}%

\begin{document}

\title{TaQ-DiT: Time-aware Quantization for Diffusion Transformers}

\author{Xinyan Liu, Huihong Shi, Yang Xu, and Zhongfeng Wang,~\IEEEmembership{Fellow,~IEEE,}
\thanks{DOI: 10.1109/TCSVT.2026.3652275}
\thanks{This work was supported by the National Key R\&D Program of China under Grant 2022YFB4400600}
\thanks{Xinyan Liu and Huihong Shi contributed equally to this work. Xinyan Liu, Huihong Shi, and Yang Xu are with the School of Electronic Science and Engineering, Nanjing University, Nanjing, China (e-mail: \{{xinyanliu, shihh, xyang}\}@smail.nju.edu.cn).}
\thanks{Zhongfeng Wang is with the School of Electronic Science and Engineering, Nanjing University, and the School of Integrated Circuits, Sun Yat-sen University (email: zfwang@nju.edu.cn).}
\thanks{Correspondence should be addressed to Zhongfeng Wang.}}

\markboth{Journal of \LaTeX\ Class Files,~Vol.~14, No.~8, August~2021}%
{Shell \MakeLowercase{\textit{et al.}}: A Sample Article Using IEEEtran.cls for IEEE Journals}

\IEEEpubid{\begin{minipage}{\textwidth}
\centering
Copyright © 2026 IEEE. Personal use of this material is permitted.\\ However, permission to use this material for any other purposes must be obtained from the IEEE by sending an email to pubs-permissions@ieee.org.
\end{minipage}}

\maketitle

\begin{abstract}
Transformer-based diffusion models, dubbed Diffusion Transformers (DiTs), have achieved state-of-the-art performance in image and video generation tasks. However, their large model size and slow inference speed limit their practical applications, calling for model compression methods such as quantization. 
Unfortunately, existing DiT quantization methods overlook (1) the impact of reconstruction\lxyRR{, a widely used method to calibrate quantization parameters,} and (2)
the varying quantization sensitivities across different layers, which hinder their achievable performance.
To tackle these issues, we propose innovative time-aware quantization for DiTs (TaQ-DiT). Specifically, (1) we observe a non-convergence issue when reconstructing weights and activations separately during quantization and introduce a joint reconstruction method to resolve this problem. (2) We discover that Post-GELU activations are particularly sensitive to quantization due to their significant variability across different denoising steps as well as extreme asymmetries and variations within each step.
To address this, we propose time-variance-aware \lxyRR{static} transformations 
to facilitate more effective quantization.
Experimental results show that when quantizing DiTs' weights to 4-bit and activations to 8-bit (W4A8), our method significantly surpasses previous quantization methods. Codes are available at \url{https://github.com/6xy-liu/Taq-DiT.git}.
\end{abstract}


\vspace{-1.2em}
\section{Introduction}
\IEEEPARstart{R}{ecently}, 
inspired by the success of Transformers \cite{ViT}, Transformer-based Diffusion Models (DMs) dubbed Diffusion Transformers (DiTs) have emerged \cite{DiT}, demonstrating superior scalability on complex generation tasks compared to U-Net-based DMs \cite{stable_diffusion}. 
However, their large model size and intensive computations involved in the iterative denoising process result in slow inference speeds \cite{PTQ4DiT}, calling for effective model compression methods.
\begin{figure}[!t]
\centering
\setlength{\abovecaptionskip}{0.1cm}
\includegraphics[width=1.0\columnwidth]{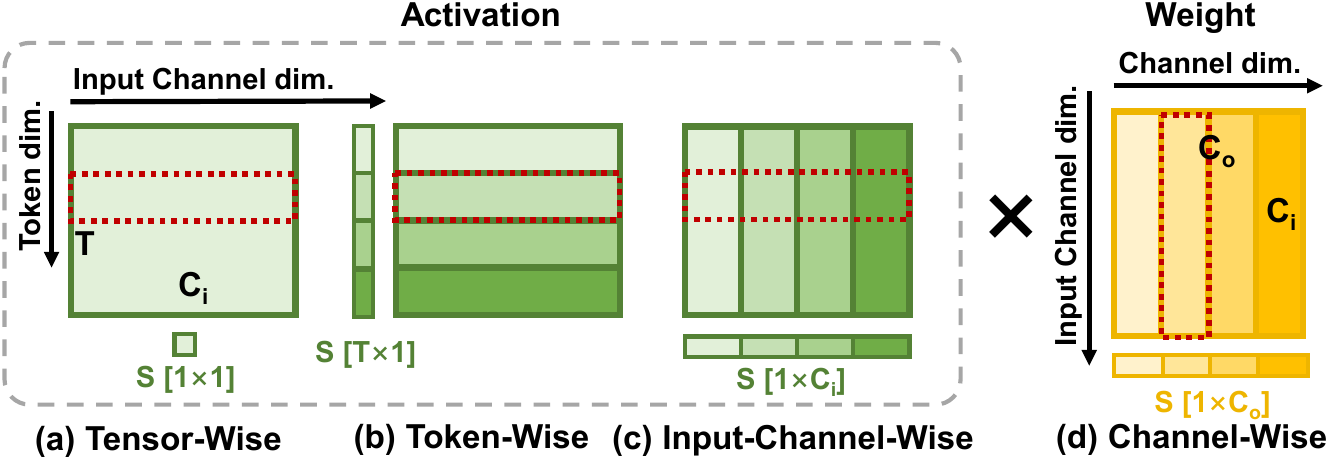}
\vspace{-1.5em}
\caption{Illustrating (a) tensor-wise, (b) token-wise, and (c) input-channel-wise quantization for activations, and (d) channel-wise quantization for weights. \lxyRR{To ensure hardware-efficient computation, we adopt channel-wise quantization for all activations and tensor-wise quantization for all weights.}}
\label{uniform_quant} \vspace{-2.5em} 
\end{figure}

\lxyRR{
Recently, many works \cite {tcsvt_dm_speedup1,tcsvt_dm_speedup2,tcsvt_dm_speedup3} have explored DiT acceleration.
Among them, quantization \cite{ptq4vit,mixedprecision}, which converts full-precision (FP) parameters into integers, is one of the most effective compression techniques. Post-Training Quantization (PTQ) is particularly attractive as it requires only a small calibration dataset and avoids costly fine-tuning \cite{zeroquant,tcsvt_BinaryViT}.}
However, existing quantization methods are primarily designed for U-Net-based DMs \cite{efficientdm,ptq4dm,ptqd,q-diffusion}, yielding performance drops when directly applied to DiTs due to their unique algorithmic properties and architectures \cite{PTQ4DiT, DiTAS}. This underscores the need for DiT-tailored quantization\lxyMinor{\cite{Q-DiT}}.
For example, PTQ4DiT \cite{PTQ4DiT} identifies the presence of salient channels that exhibit extreme magnitudes and variability over timesteps, and proposes dedicated techniques to facilitate DiT quantization.
\lxyRR{Additionally, DiTAS \cite{DiTAS} proposes temporal-aggregated smoothing for channel-wise outlier mitigation and a layer-wise grid search to improve quantization performance.}
Despite the effectiveness of these methods, they overlook 
(1) the impact of reconstruction on DiT quantization and (2) the distinct quantization sensitivities of different layers within DiTs.

To push the frontier of DiT quantization, we propose TaQ-DiT and make the following contributions:

\vspace{-0.2em}
\begin{itemize}
\item We observe that the widely adopted reconstruction method, which optimizes the quantization of weights and activations separately, suffers from non-convergence issues. To address this, we propose using \textbf{\textit{joint reconstruction}} that integrates both to improve compatibility.
\item 
We further identify that activations after GELU (\textbf{\textit{Post-GELU activations}}) within DiTs are particularly sensitive to quantization, due to their (1) significant variability across different denoising steps and (2) extreme asymmetries and variations along the input channel dimension within each step. To address them, we propose \textbf{time-variance-aware \lxyRR{static} transformations} that integrate innovative \textbf{\textit{momentum-based shifting}} and \textbf{\textit{reconstruction-driven migration}} techniques to facilitate quantization.
\item Extensive experiments demonstrate that TaQ-DiT outperforms the SOTA DiT-dedicated quantization methods by $\downarrow$$1.09$ FID (lower is better) when weights are quantized to $4$-bit and activations to $8$-bit (W4A8). 
\end{itemize}

\IEEEpubidadjcol
\vspace{-1.0em}
\section{Preliminaries}
\label{sec:Preliminaries}
\subsection{The Structure of DiT}

\lxyRR{DiT is a Transformer-based DM.
Following the basic principles of DMs\cite{DDPM}, timesteps index the forward/reverse noise processes to indicate progression. The forward process of a DM progressively adds noise to $x_0$ over timesteps, where the conditional distribution at each step is given by $\label{q}
q(x_t | x_0) = \mathcal{N}\left(x_t; \sqrt{\bar{\alpha}_t} x_0, (1 - \bar{\alpha}_t)\mathbf{I}\right)$, }
\lxyRR{where ${\bar{\alpha}_t}$ are predefined hyperparameters. By reparameterization, $x_t$ can be directly sampled as $x_t = \sqrt{\bar{\alpha}_t} x_0 + \sqrt{1 - \bar{\alpha}_t} \epsilon_t, \epsilon_t \sim \mathcal{N}(0, \mathbf{I})$.}
\lxyRR{In the reverse denoising process, DMs are trained to progressively generate images from noise $x_T$ through neural networks such as U-Nets or Transformers, which predict the statistics of $p_{\theta}$: }

\vspace{-0.5em}
\begin{equation}\label{p}
p_{\theta}(x_{t-1} | x_t) = \mathcal{N}\left(\mu_{\theta}(x_t), \Sigma_{\theta}(x_t) \right).
\end{equation}
\vspace{-1.em}

\lxyRR{Thus, by initializing with random noise, we can generate images through a step-by-step sampling process $x_{t-1} \sim p_{\theta}(x_{t-1} \mid x_t)$ over $T$ timesteps\cite{DiT}.}

\lxyRR{The Transformer-based DiT backbone comprises two conditioned components receiving class and timestep information during the denoising process: Multi-Head Self-Attention (MHA) layers and Pointwise Feedforward (PF) networks with the GELU activation functions.}
Therefore, considering both the time-dependent characteristic and the architectural property of Transformer blocks is essential for designing an effective quantization scheme for DiT.

\vspace{-1.2em}
\subsection{Uniform Quantization}
Uniform quantization is the most common and easily implementable model compression method, which converts the floating-point parameters $X$ into $b$-bit integer $Q(X)$ via:
\begin{equation}\label{uniformquant}
Q(X)=\operatorname{clip}\left(\left\lfloor{{X/}{S}}\right\rceil+Z, 0, 2^{b}-1\right),
\end{equation}
where the $S$ and $Z$ represent the scaling factor and zero point, respectively, detailed by:

\vspace{-1.2em}
\begin{equation}\label{scale_and_zaro-point}
\small
S=\frac{\operatorname{max}(X)-\operatorname{min}(X)}{2^{b}-1}, \
Z=\operatorname{clip}\left(\left\lfloor-\frac{\operatorname{min}(X)}{S}\right\rceil, 0, 2^{b}-1\right). 
\end{equation}

Uniform quantization can be applied at different levels of granularity. Specifically, as shown in Fig. \ref{uniform_quant}(a), activations typically use tensor-wise quantization \cite{DiTAS} with a single scaling factor for implementation ease. For finer granularity in Fig. \ref{uniform_quant}(b), as elements within each activation token need to be multiplied and summed with elements in the corresponding weight channel, token-wise quantization\cite{zeroquant} assigns distinct scaling factors to individual activation tokens, improving accuracy without compromising hardware efficiency. Similarly, weights often employ channel-wise quantization \cite{li2023repq} with per-channel scaling factors (Fig. \ref{uniform_quant}(d)).

\lxyMajor{Reconstruction \cite{brecq,qdrop,tcsvt_recon} improves the performance of low-bit quantization by optimizing quantization parameters, such as scaling factors and zero points, through minimizing the discrepancy between the outputs of quantized and full-precision models using a small calibration dataset and second-order loss analysis. This approach effectively preserves model performance without full retraining \cite{ptq4vit,q-diffusion} and is widely used in PTQ.}
\section{Methodology}
\subsection{Joint Reconstruction for DiT Quantization} \label{sec:Reconstruction method selection}
\emph{\textbf{Observation: Non-Convergence Issue in Quantization Reconstruction for Activations.}} 
To explore the challenges and opportunities in DiT quantization, we start by applying existing DM quantization approaches \cite{q-diffusion,ptq4vit}, which reconstruct scaling factors for weights and activations separately following \cite{brecq}. However, as listed in Table \ref{tab:reconstruction_method},
we observe that quantizing weights to $4$-bit (W4) alone leads to a $\uparrow$$\mathbf{3.01}$ Fr{\'{e}}chet Inception Distance (FID), while further quantizing activations to $8$-bit (W4A8) results in a significant $\uparrow$$\mathbf{7.97}$ FID. This indicates that this widely adopted reconstruction method is unsuitable for DiT quantization, particularly for activations.
We attribute this failure to the non-convergence issue during reconstruction. As depicted in Figs. \ref{timestep_loss}(a) and \ref{timestep_loss}(b), although quantization reconstruction for weights converges, \lxyRR{the reconstruction for activations exhibits oscillation}, thereby limiting the achievable performance in DiT quantization.
\begin{table}[t]
    \centering
    \renewcommand{\arraystretch}{1.2}
    \caption{Comparisons of different reconstruction methods on DiT \label{tab:reconstruction_method}} 
    \vspace{-0.7em}
    \setlength{\tabcolsep}{0.95em}
    \resizebox{0.9\linewidth}{!}{
    \begin{threeparttable}
        \begin{tabular}{cc|cccc}
        \Xhline{3\arrayrulewidth}
            \multicolumn{2}{c|}{\textbf{Method\textsuperscript{1}}}               & \textbf{FID$\downarrow$}    & \textbf{sFID$\downarrow$}   & \textbf{IS$\uparrow$}       & \textbf{Precision$\uparrow$} \\ \hline \hline
            \multicolumn{2}{c|}{\textbf{Full Precision\cite{PTQ4DiT}}} & 5.00   & 19.02  & 274.78   & 0.8149   \\ \hline
            \multicolumn{1}{c|}{\multirow{2}{*}{\textbf{Separate}}} & \textbf{W4}       & 8.01   & 20.10  & 183.97   & 0.7302   \\ 
            \multicolumn{1}{c|}{}                  & \textbf{W4A8} & 12.97  & 25.20  & 147.58   & 0.6844   \\ \hline
            \multicolumn{1}{c|}{\multirow{2}{*}{\textbf{Joint}}}    & \textbf{W4\textsuperscript{2}}    & 6.12  & 18.92  & 236.61  & 0.7857 \\  
            \multicolumn{1}{c|}{}                  & \textbf{W4A8} & 6.61  & 19.50  & 231.60   & 0.7713\\ \Xhline{3\arrayrulewidth}
        \end{tabular}
            \begin{tablenotes}
                \footnotesize
                \item[1] 
                Tested on 10,000 samples generated by 100 timesteps. \lxyRR{We only quantize MHA and PF layers
                \item[2] Measured after activation and weights are jointly reconstructed 
                }
            \end{tablenotes} 
    \end{threeparttable}} \label{tab:reconstruction}
    \vspace{-2.0em}
\end{table}

\emph{\textbf{Method:
Joint Reconstruction of Both Activations and Weights.}}
To address this issue and boost quantization performance, motivated by QDrop \cite{qdrop}, we integrate the reconstruction of weights and activations to facilitate their compatibility:
\begin{equation}\label{recon}
\min_{S_W, S_A} \mathcal{L} \left[ B(Q(W), Q(A)) - B(W, A) \right],
\end{equation}
where $B$ represents the blocks or layers to be reconstructed, and $Q(\cdot)$ \lxyRR{denotes} the quantization function. $W$ and $A$ represent weights and activations, respectively, with $S_W$ and $S_A$ as their corresponding scaling factors.
\lxyRR {Eq. (\ref{recon}) calibrates $S_W$ and $S_A$ to minimize the reconstruction loss measured by the Mean Squared Error (MSE), which means the output gap between the jointly quantized and full-precision models\cite{brecq,qdrop}.}
As shown in Fig. \ref{timestep_loss}(c) and Table \ref{tab:reconstruction_method}, the joint reconstruction resolves non-convergence and improves quantization performance by $\downarrow$$\mathbf{6.36}$ FID on W4A8 compared to the separate reconstruction method. \lxyRR{Moreover, joint reconstruction improves W4 inference by $\downarrow$$\mathbf{1.89}$ FID compared to the baseline, revealing how activation reconstruction enhances weight optimization.}
\vspace{-1em}
\subsection{Time-Variance-aware Static Transformation for Post-GELU Activations} \label{sec:Shifting and Migration}
As demonstrated in Table \ref{tab:reconstruction_method}, despite the effectiveness of joint reconstruction, there still exists a non-negligible performance gap ($\uparrow$$\mathbf{1.61}$ FID) between the W4A8 quantized model and the full-precision counterpart.
To investigate the reason for this, we further conduct ablation studies on the quantization of different layers. As seen in Table \ref{tab:GELU_sensitivity}, PF layers are more sensitive to quantization than MHA blocks, leading to even worse performance ($\uparrow$$\mathbf{2.66}$ FID) than quantizing the entire model. Going a step further, we identify that activations after GELU (\textbf{Post-GELU activations}) are the primary contributors to the performance drop in PFs by $\uparrow$$\mathbf{1.47}$ FID. 
\begin{table}[t]
    \centering
    \setlength{\tabcolsep}{1.8pt}
    \renewcommand{\arraystretch}{1.2}
    \caption{Impact of W4A8 quantization on different layers within DiT \label{tab:GELU_sensitivity}} 
    \vspace{-0.8em}
    \resizebox{0.93\linewidth}{!}{
    \begin{threeparttable}
        \begin{tabular}{c|cccc}  \Xhline{3\arrayrulewidth}
            \textbf{Method*}  & \textbf{FID$\downarrow$} & \textbf{sFID$\downarrow$} & \textbf{IS$\uparrow$} & \textbf{Precision$\uparrow$} \\ 
            \hline \hline
            \textbf{Full Precision\cite{PTQ4DiT}}  & 5.00  & 19.02 &  274.78  &  0.8149        \\
            \hline
            \textbf{\lxyRR{MHAs + PFs}}       & 6.61  & 19.50   &  231.60  & 0.7713 \\ 
            \textbf{MHAs} & 5.74 & 18.67    &  260.87   & 0.8066 \\ 
            \textbf{\lxy{PF}s}       & 7.66 & 20.69    &  208.73   & 0.7497 \\
            \textbf{\lxy{PF}s except Post-GELU Activations}       & 6.19 & 19.15    &  234.48   & 0.7779 \\
            \Xhline{3\arrayrulewidth}
        \end{tabular}
        \begin{tablenotes}
            \footnotesize
            \item[*] Tested on 10,000 samples generated by 100 timesteps.
        \end{tablenotes} 
    \end{threeparttable}}
    \vspace{-1.0em}
\end{table}

To address this, we first visualize and analyze Post-GELU activations, then introduce our \lxyRR{\textbf{time-variance-aware static transformation} to facilitate quantization, integrating momentum-based shifting (Method 1) and reconstruction-driven migration (Method 2), which are applied on top of the tensor-wise activation quantization.}

\begin{figure}[t]
\centering
\setlength{\abovecaptionskip}{0.1cm}
\includegraphics[width=0.92\columnwidth]{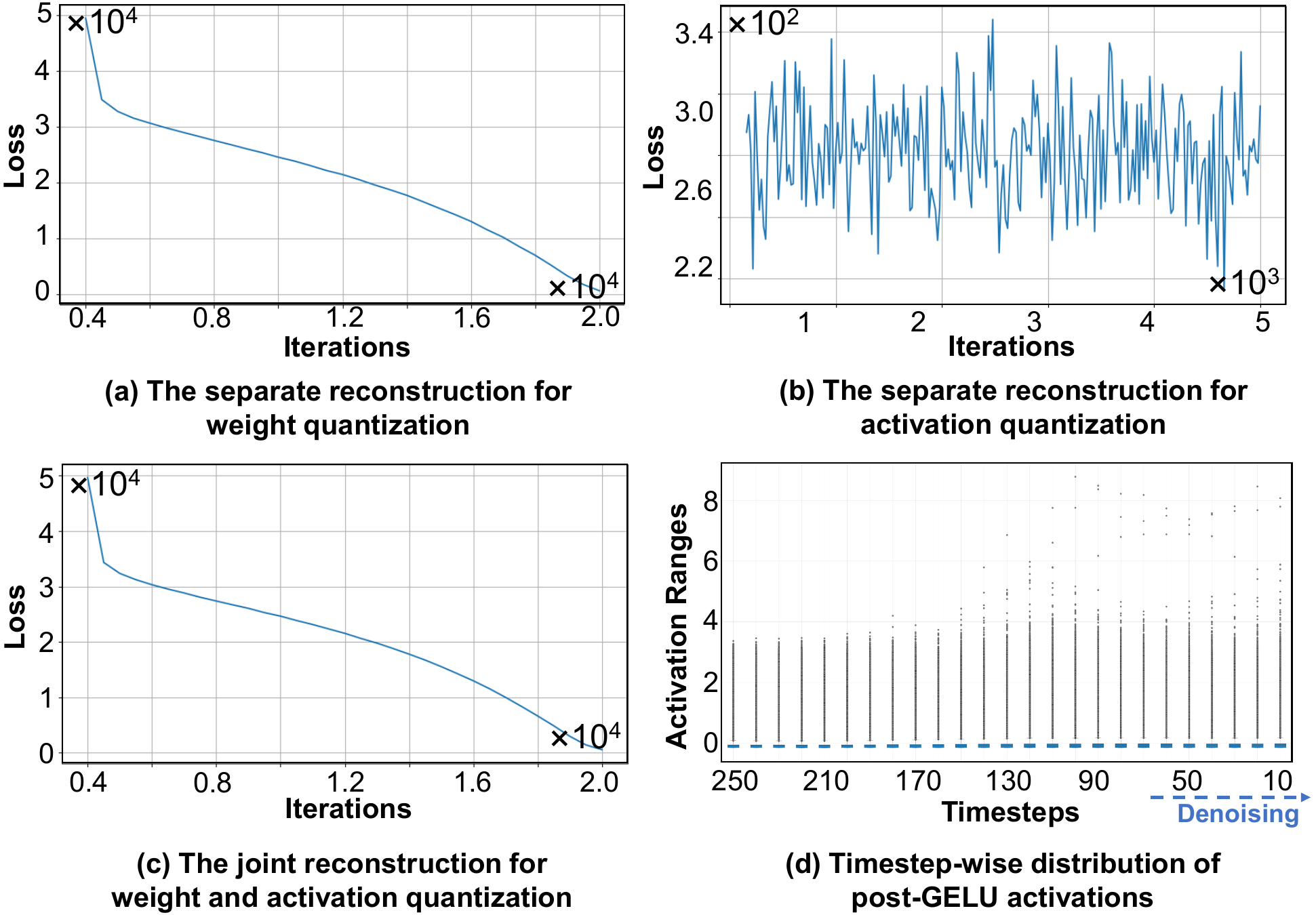}
\caption{(a)-(c) The trajectories of separate and joint reconstruction.
(d) Activation ranges of Post-GELU activations across different timesteps. }
\label{timestep_loss} \vspace{-1.2em} 
\end{figure}

\emph{\textbf{Observation 1: Significant Variability Across Different Denoising Steps.}}
As shown in Figs. \ref{timestep_loss}(d), Post-GELU activations exhibit significant variability across different timesteps during denoising, posing challenges to quantization. To accommodate timestep dependency, two main quantization methods can be considered. The first is dynamic quantization\cite{smoothquant,zeroquant}, which computes unique scaling factors for each timestep during inference, enhancing performance but increasing latency. The second is static quantization \cite{smoothquant,PTQ4DiT}, which relies on aggregated statistics across all timesteps to precompute a single scaling factor during calibration, improving quantization efficiency but often leading to performance degradation.

\begin{figure}[t]
\centering
\setlength{\abovecaptionskip}{0.1cm}
\includegraphics[width=1.0\columnwidth]{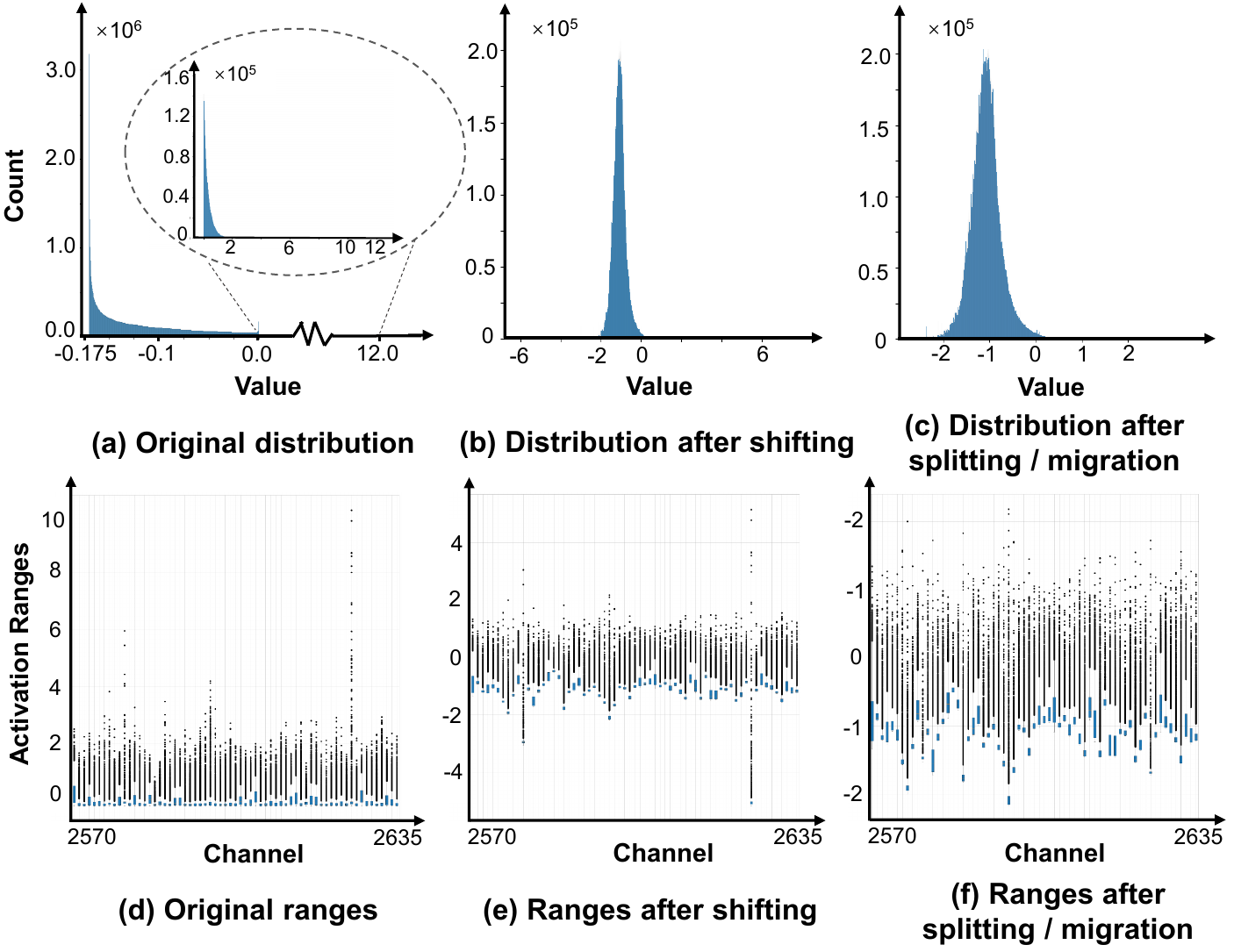}
\vspace{-2em}
\caption{Distributions and channel-wise ranges of Post-GELU activations, where (a)(d) original, (b)(e) after shifting, and (c)(f) after splitting/migration.}
\label{distrib_for_2.1} \vspace{-1em}
\end{figure}

\lxyMajor{\emph{\textbf{Observation 2: Extreme Asymmetries Within Each Denosing Step.}}}
\shh{
Besides step-wise variability, as shown in Figs. \ref{distrib_for_2.1}(a) and \ref{distrib_for_2.1}(d), Post-GELU activations exhibit extreme irregularities and asymmetries within each denoising step, with a clear boundary between positive and negative values. These irregularities result in the majority of activations having extremely low quantization resolution, thus challenging the vanilla uniform quantization and limiting its achievable performance. For instance, for the Post-GELU activation of the $20^{th}$ block at the $60^{th}$ timestep, when quantizing it to $8$-bit, negative values that constitute $79.7\%$ of the activations can be allocated merely $1.6\%$ of the quantization bins. Additionally, although positive activations are assigned the majority of quantization bins ($98.4\%$), due to the existence of extremely large values, \lxy{$99.6\%$} of positive values only have \lxy{$16.0\%$} quantization bins.}

\emph{\textbf{Method 1: Momentum-Based Shifting for Extreme Asymmetries.}}
\label{sec: shifting}
\shh{To regularize distributions of Post-GELU activations and facilitate quantization, we propose to adopt channel-wise shifting to symmetrize their distributions before quantization. Specifically, we subtract activation elements $a^i$ in $i^{th}$ channel by its channel-wise mean (shifting) value $v^i$ to obtain the symmetrized channel $\widetilde{a}^i$ for easing quantization:}
\begin{equation}
\widetilde{a}^i =a^i-v^i, \ \text{where} \label{eq:shifting_activation}
\end{equation}
\begin{equation}
v^i=(\operatorname{max}(a^i)+\operatorname{min}(a^i))/{2}.  \label{eq:shiting_value}
\end{equation}
\shh{To preserve mathematical equivalence while avoiding online computations, the shifting values $V$ can be precomputed with weights $W$ and then merged into the original bias $b$ to form a new bias $\widetilde{b}$, where $A$ and $\widetilde{A}$ represent the original and shifted activations, respectively:}
\begin{equation}
AW+b=(\widetilde{A}+V)W+b=\widetilde{A}W+(VW+b)=\widetilde{A}W+\widetilde{b},
\end{equation}

\shh{However, as shown in Figs. \ref{timestep_loss}(d) and explained in \textbf{\textit{Observation 1}}, activation ranges vary across timesteps, necessitating \lxyRR{\textbf{dynamic shifting values} to accommodate temporal variability. However, this yields online calculations of shifting values for each denoising step during inference, leading to huge computational and latency overheads.} Thus, we propose a momentum-based method to pre-compute shifting values during calibration\lxyRR{, ensuring they remain \textbf{static} during inference}:}
\begin{equation}
\vspace{-0.15em}
v_{\text{updated}}=\beta v_{\text{previous}}+(1-\beta)v_{\text{current}}, \label{eq:shiting_update}
\vspace{-0.15em}
\end{equation}
\shh{where $\beta$ is the momentum coefficient and is set to $0.95$ following \cite{zeroquant}. $v_{\text{current}}$ represents the shifting value calculated from the calibration data at the current timestep following Eq. (\ref{eq:shiting_value}). Meanwhile, $v_{\text{updated}}$ and $v_{\text{previous}}$ donate the updated and previous shifting values, respectively.
This implies that rather than using Eq. (\ref{eq:shiting_value}) to dynamically calculate $v^i$ for shifting activations in Eq. (\ref{eq:shifting_activation}), we use Eq. (\ref{eq:shiting_update}) instead to accommodate temporal variability and enhance quantization performance.}

\shh{As depicted in Figs. \ref{distrib_for_2.1}(b), the shifted activations transfer from power-law-like distributions to more regular and symmetric Gaussian distributions. 
For example, in the \textit{shifted} Post-GELU activation of the $20^{th}$ block at the $60^{th}$ timestep, when quantized to $8$-bit, $79\%$ of the values can now be allocated to $6.8\%$ of the quantization bins, which is $\mathbf{4.25}\times$ larger than before, demonstrating our effectiveness. }

\emph{\textbf{Observation 3: Channel-Wise Variations of Post-GELU Activations.}}
\label{sec: spliting}
\shh{Although our proposed momentum-based shifting method helps regularize the distributions of Post-GELU activations, the existence of outliers (a small number of extremely large values) still leads to low quantization resolution for the majority of values.
To address this, we take a step further to observe and analyze the distributions of outliers. As shown in Figs. \ref{distrib_for_2.1}(e), outliers are concentrated in only a few \textit{channels}, yielding significant variations along the channel dimension.
A natural solution is to adopt channel-wise quantization, which assigns distinct scaling factors to each channel to enhance quantization performance, as shown in Figs. \ref{uniform_quant}(c). However, as discussed in \lxy{Sec. \ref{sec:Preliminaries}-B}, this leads to floating-point summations among channels, which hinders quantization efficiency.}

\emph{\textbf{Method 2: Reconstruction-Driven \lxy{Migration to Remove Outliers}.}}
\lxyMinor{To handle channel-wise variations, rather than adopting channel-wise quantization, we propose applying channel-wise \textbf{\textit{outlier factors}} on top of \textit{tensor-wise} quantization \textit{exclusively} for outlier channels, thereby enhancing quantization efficiency.} Specifically, after shifting input activations $A$, we identify the top-k channels with the largest ranges of shifted activations $\widetilde{A}$ as outlier channels $\widetilde{a}_{o}^{\text{top-k}}$, while treating the remaining channels as normal ones $\widetilde{a}_{n}$. Then, the tensor-wise scaling factor is computed based on $\widetilde{a}_{n}$ following Eq. (\ref{scale_and_zaro-point}). Subsequently, the channel-wise \lxyMinor{outlier factor} $m^i$ for the $i^{th}$ outlier channel $\widetilde{a}_{o}^{i}$ can be calculated as follows:
\vspace{-0.4em}
\begin{equation}\label{migration-factor}
m^i=\operatorname{round}(\frac{\text{max}(\widetilde{a}_{\text{o}}^i)}{\text{max}(\widetilde{a}_{\text{n}})}).
\vspace{-0.4em}
\end{equation}
\lxyRR{The most straightforward approach is to employ \textbf{dynamic splitting}, which simultaneously extracts quantization parameters, top-k outliers, and their corresponding outlier factors while dividing $\widetilde{a}_{o}$ by $m$ into multiple sub-channels to mitigate outliers during inference for timestep-wise variability.}
As shown in Figs. \ref{distrib_for_2.1}(c) and \ref{distrib_for_2.1}(f), this approach significantly reduces activation ranges, thus enhancing quantization resolution and performance. However, to maintain the same functionality, the matrix sizes of both activations and weights must be expanded, leading to computational and memory overhead.

\lxyRR{For hardware efficiency, we statistically analyze top-k outliers across timesteps offline \lxyMinor{based on a small dataset, which may slightly differ from real inference but remain generally acceptable\cite{brecq,q-diffusion},} and apply channel-wise migration accordingly, which migrates outliers in $\widetilde{a}_{o}$ into corresponding {rows} in the subsequent weights via \lxyMinor{outlier factors} $m$  as follows:}
\vspace{-0.1em}
\begin{align}\label{migration}
    &\begin{bmatrix}
      \widetilde{a}^1_n \cdots \widetilde{a}_o^i\cdots  \widetilde{a}^{C_i}_n
    \end{bmatrix}
    \times
    \begin{bmatrix}
     w^1 \cdots w^i \cdots w^{C_i}
    \end{bmatrix}^\top \nonumber \\
    = & \begin{bmatrix}
      \widetilde{a}^1_n \cdots \frac{\widetilde{a}_o^i}{m^i} \cdots  \widetilde{a}^{C_i}_n
    \end{bmatrix}
    \times
    \begin{bmatrix}
     w^1 \cdots w^i\times m^i \cdots w^{C_i}
    \end{bmatrix}^\top,
\end{align}
\vspace{-0.1em}
\lxy{where ${\widetilde{a}^{i}}$$\in $$\mathbb{R}^T$ is the $i^{th}$ column/channel of shifted activations and ${w^{i}}$$\in$$\mathbb{R}^{C_o}$is the $i^{th}$ row/input channel of weights.
$C_i$, $C_o$, and $T$ are the numbers of input channels, output channels, and tokens, respectively.} 
\lxyRR{To deal with step-wise variability while maintaining quantization efficiency, we optimize outlier factors only once during the joint reconstruction by the \textbf{static migration}. Specifically, we first initialize them using Eq. (\ref{migration-factor}) as a good starting point, then fine-tune them along with scaling factors using our proposed joint reconstruction to boost performance.
This approach achieves the same effect as dynamic splitting in eliminating channel-wise outliers, while avoiding matrix expansion and the associated overheads.} 
\lxyRR{While hardware-efficient, it could impair weight quantization. However, experiments confirm weight robustness (Sec. \ref{sec:ablation_study}, Table \ref{tab:parameters}), justifying our channel-wise migration choice.}
\vspace{-0.5em}

\section{Experiment}
\subsection{Experimental Settings}

\shh{To ensure fair comparisons, we adopt similar experimental settings to prior DiT quantization works \cite{DiTAS,PTQ4DiT}. \underline{\textbf{\textit{Dataset}}}: We evaluate TaQ-DiT on the ImageNet dataset at a resolution of $256$$\times$$256$ \lxyRR{with the DDPM sampler}. 
For the generation process, we set the sampling steps to $100$ and use a Classifier-Free Guidance (CFG) score of $1.50$.
\lxyRR{To construct our calibration dataset, we uniformly select $25$ steps from the total steps and generate $256$ samples at each selected step, then randomly shuffle them across the chosen steps.}
\underline{\textbf{\textit{Quantization Settings}}}: We utilize uniform quantization for both activations and weights, \lxyRR{with $4$-bit channel-wise quantization for weights and $8$-bit tensor-wise quantization for activations.}
Our codes are implemented on PyTorch and experimented on NVIDIA A100 GPUs.
\underline{\textbf{\textit{Baselines}}}: We compare with \textbf{\textit{six}} baselines, including (i) PTQ works tailored for U-Net-based DMs, i.e., PTQ4DM \cite{ptq4dm}, Q-Diffusion \cite{q-diffusion}, PTQD \cite{ptqd}, and RepQ \cite{li2023repq}, which are applied to DiTs following quantization settings in PTQ4DiT \cite{PTQ4DiT} for fair comparisons, and (ii) PTQ works dedicated to DiTs, i.e., PTQ4DiT\cite{PTQ4DiT} and DiTAS\cite{DiTAS}.
\underline{\textbf{\textit{Metrics}}}: To quantitatively evaluate generated images, we sample $10,000$ images and assess them using \textbf{\textit{four}} metrics:  Fr{\'{e}}chet Inception Distance (FID), spatial FID (sFID), Inception Score (IS), and Precision.
}

\vspace{-2.3em}
\subsection{Comparisons With SOTA Works}
\begin{table}[t]
    \centering
    \renewcommand{\arraystretch}{1.2}
    \caption{Quantization Performance Comparisons on DiT \label{tab:compare}} 
    \vspace{-0.8em}
    \setlength{\tabcolsep}{0.6em}
    \resizebox{0.95\linewidth}{!}{
    \begin{threeparttable}
        \begin{tabular}{cl|ccccc} 
            \Xhline{3\arrayrulewidth}
            \multicolumn{2}{c|}{\textbf{Method}}  & \textbf{FID$\downarrow$} & \textbf{sFID$\downarrow$} & \textbf{IS$\uparrow$} & \textbf{Precision$\uparrow$} \\ 
            \hline \hline
            \rowcolor{gray!15}\multicolumn{2}{c|}{\textbf{Full Precision*}}    & 5.00                     & 19.02                     &  274.78               &  0.8149       \\
            \hline
            \multicolumn{1}{c|}{\multirow{6}{*}{\textbf{All layers}}} & {PTQ4DM* \cite{ptq4dm}}            & 89.78  & 57.20    &  26.02   & 0.2146 \\ 
            \multicolumn{1}{c|}{} &{Q-Diffusion* \cite{q-diffusion}}        & 54.95  & 36.13    &  42.80   & 0.3846 \\ 
            \multicolumn{1}{c|}{} &{PTQD* \cite{ptqd}}               & 55.96  & 37.24    &  42.87   & 0.3948 \\
            \multicolumn{1}{c|}{} &{RepQ* \cite{li2023repq}}               & 26.64  & 29.42    &  91.39   & 0.4347 \\
            \multicolumn{1}{c|}{} &{PTQ4DiT* \cite{PTQ4DiT}}            & 7.75  & 22.01    &  190.38   & 0.7292 \\
            \multicolumn{1}{c|}{} &\ \cellcolor{dark-green!15}\textbf{Ours}               & \cellcolor{dark-green!15}\textbf{6.66}     & \cellcolor{dark-green!15}\textbf{19.27}    &  \cellcolor{dark-green!15}\textbf{221.61}   & \cellcolor{dark-green!15}\textbf{0.7781}\\
            \hline
            \multicolumn{1}{c|}{\multirow{2}{*}{\textbf{MHAs + PFs}}} & {DiTAS \cite{DiTAS}}              & 6.86  & 19.64    &  218.04   & 0.7638 \\
            \multicolumn{1}{c|}{} &\cellcolor{dark-green!15}\textbf{Ours}               & \cellcolor{dark-green!15}\textbf{6.03}     & \cellcolor{dark-green!15}\textbf{18.50}    &  \cellcolor{dark-green!15}\textbf{232.36}   & \cellcolor{dark-green!15}\textbf{0.7769}\\
            \Xhline{3\arrayrulewidth}
        \end{tabular}
        \begin{tablenotes}
            \footnotesize
            \item[*] Results are derived from PTQ4DiT.
        \end{tablenotes} 
    \end{threeparttable}}
 \vspace{-2.0em}
\end{table}

\begin{figure}[t]
\centering
\setlength{\abovecaptionskip}{0.1cm}
\includegraphics[width=0.8\columnwidth]{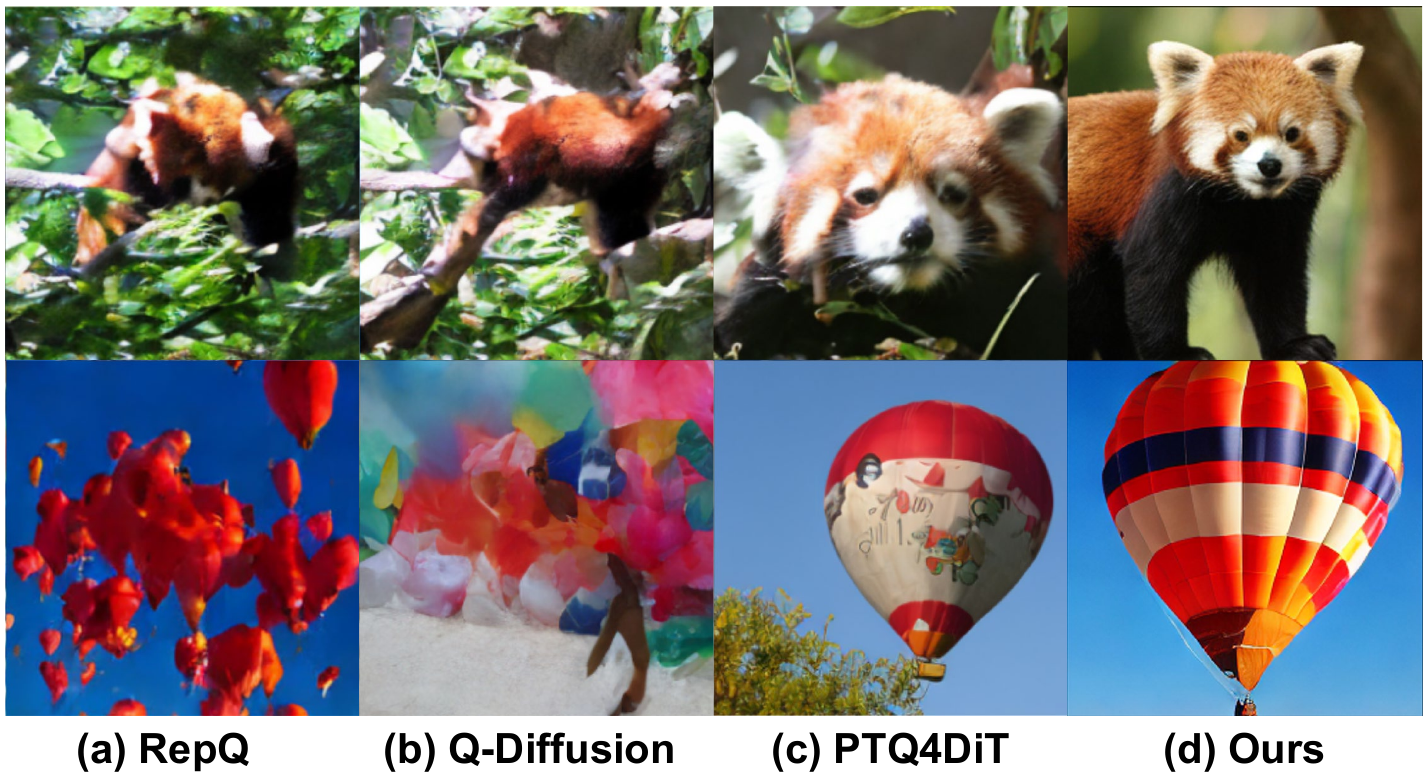}
\vspace{-0.4em}
\caption{Comparisons of generated images between ours and baselines.}
\label{visual} 
\vspace{-0.9em}
\end{figure}

As shown in Table \ref{tab:compare}, TaQ-DiT surpasses all baselines, validating our effectiveness.
\lxyRR{Specifically, with all layers quantized following the setting of PTQ4DiT\cite{PTQ4DiT}: \textbf{(1)} when compared with quantization methods tailored for U-Net-based DMs, TaQ-DiT outperforms them by a huge margin, e.g., $\downarrow$$49.30$$\sim$$\downarrow$$83.12$ FID. This highlights the need for DiT-specialized quantization. \textbf{(2)} TaQ-DiT also exhibits superior results compared to DiT-dedicated baseline PTQ4DiT\cite{PTQ4DiT} with $\downarrow$$1.09$ and $\uparrow$$31.23$ on FID and IS, respectively. Notably, when quantizing only MHA and PF layers, our method still maintains superior results ($\downarrow$$0.83$ FID and $\uparrow$$14.32$ IS) compared to DiTAS\cite{DiTAS} under the same setting.} We attribute the improvements to (i) the joint reconstruction to boost overall performance and (ii) our time-variance-aware static transformation to facilitate quantization for sensitive Post-GELU activations.
\shh{Besides, as shown in Fig. \ref{visual}, we achieve much better visual quality, further highlighting our superiority from a qualitative perspective.}

\vspace{-1.0em}
\subsection{Ablation Study}
\label{sec:ablation_study}
\begin{table}[t!]
    \centering
    \renewcommand{\arraystretch}{1.2}
    \caption{The effect of different components proposed in the paper \label{tab:ablation}} 
    \vspace{-0.5em}
    \setlength{\tabcolsep}{0.3em}
    \resizebox{0.85\linewidth}{!}{
    \begin{threeparttable}
        \begin{tabular}{cc|cccc} 
            \Xhline{2\arrayrulewidth}
            \multicolumn{2}{c|}{\textbf{Method}}  & \multirow{3}{*}{\textbf{FID$\downarrow$}} & \multirow{3}{*}{\textbf{sFID$\downarrow$}} & \multirow{3}{*}{\textbf{IS$\uparrow$}} & \multirow{3}{*}{\textbf{Precision$\uparrow$}} \\ \cline{1-2}
            {\textbf{Joint \  }}  &\textbf{\lxyRR{Time-Variance-Aware}}  &  & \\
           {\textbf{Reconstruction \  }}  &\textbf{\lxyRR{Static Transformation}} & & \\
            \hline \hline
            \rowcolor{gray!15}-    &   -   & 5.00  & 19.02 &  274.78  &  0.8149        \\
            \hline
            {\color{purple}{\xmark}}  &{\color{purple}{\xmark}}               & 12.97  & 25.20  & 147.58   & 0.6844   \\
            {\color{dark-green}{\cmark}}   &{\color{purple}{\xmark}}                & 6.61  & 19.50   &  231.60  & 0.7713 \\ 
            \rowcolor{dark-green!15}  {\color{dark-green}{\cmark}}  &{\color{dark-green}{\cmark}}       & \textbf{6.03}     & \textbf{18.50}    &  \textbf{232.36}   & \textbf{0.7769}\\
            \Xhline{3\arrayrulewidth}%
        \end{tabular}
    \end{threeparttable}}
    \vspace{-1.em}
\end{table}
\textbf{Effectiveness of Proposed Methods.} 
As shown in Table \ref{tab:ablation}, \textbf{(1)} our {{{joint reconstruction method}}} results in significant performance improvements, e.g., $\downarrow$$\mathbf{6.36}$ FID and $\uparrow$$\mathbf{84.02}$ IS. \textbf{(2)} Incorporating it with our \lxyRR{time-variance-aware static transformation} methods further enhances the performance, e.g., $\downarrow$$\mathbf{6.94}$ FID and $\uparrow$$\mathbf{84.78}$ IS, which validates our effectiveness.

\lxy{\textbf{Comparison With Dynamic Quantization.}} By integrating \textbf{\textit{momentum-base shifting}} and \textbf{\textit{reconstruction-driven migration}}, as seen in Table \ref{tab:parameters}, our \lxyRR{time-variance-aware static transformation achieves comparable performance ($\downarrow$$\mathbf{0.13}$ FID and $\downarrow$$\mathbf{2.69}$ IS) compared to the quantization-inefficient dynamic counterpart, which needs online computations for each denoising step} \lxyRR{(dynamic shifting and splitting discussed in Method 1 and Method 2 of Sec. \ref{sec:Shifting and Migration}, respectively)}.
This highlights our effectiveness in both performance and efficiency.
\begin{table}[t!]
    \centering
    \vspace{-0.5em}
    \renewcommand{\arraystretch}{1.2}
    \caption{The method to get shift and split parameters \label{tab:parameters}} 
    \vspace{-0.5em}
    \setlength{\tabcolsep}{0.6em}
    \resizebox{\linewidth}{!}{
    \begin{threeparttable}
        \begin{tabular}{c|cccc} 
            \Xhline{3\arrayrulewidth}
            \textbf{Method*}  & \textbf{FID$\downarrow$} & \textbf{sFID$\downarrow$} & \textbf{IS$\uparrow$} & \textbf{Precision$\uparrow$} \\ 
            \hline \hline
            \rowcolor{gray!15}\textbf{Full Precision\cite{PTQ4DiT}}  & 5.00  & 19.02 &  274.78  &  0.8149        \\
            \hline
            \textbf{\lxyRR{Dynamic Shifting and Splitting (Vanilla)}}                       & 6.16  & 18.74  & 235.05   & 0.7840   \\
            \rowcolor{dark-green!15} \textbf{\lxyRR{Time-Variance-Aware Static Transformation (Ours)}}           & \textbf{6.03}     & \textbf{18.50}    &  \textbf{232.36}   & \textbf{0.7769}\\ 
            \Xhline{3\arrayrulewidth}
        \end{tabular}
        \begin{tablenotes}
            \footnotesize
            \item[*] The top-2\% channels with the largest ranges of shifted activations are regarded to be split or migrated. \lxyRR{We only quantize MHA and PF layers.}
        \end{tablenotes} 
    \end{threeparttable}}
    \vspace{-1.7em}
\end{table}

\section{Conclusion}
We have proposed and validated TaQ-DiT, an advanced post-training quantization framework for DiTs. First, we propose leveraging joint reconstruction of activations and weights to resolve the non-convergence issue.  
Then, we develop innovative momentum-based shifting and reconstruction-driven migration techniques to facilitate the effective quantization of sensitive Post-GELU activations. Experimental results validate the effectiveness of our method and 
show its capability of outperforming state-of-the-art quantization methods and narrowing the gap between quantized and full-precision models.
\vspace{-2.0em}

\bibliographystyle{IEEEtran}
\bibliography{QDiT}

@INPROCEEDINGS{DiT,
  author={Peebles, William and others},
  booktitle={2023 IEEE/CVF International Conference on Computer Vision (ICCV)}, 
  title={Scalable Diffusion Models with Transformers}, 
  year={2023},
  volume={},
  number={},
  pages={4172-4182},
  doi={10.1109/ICCV51070.2023.00387}}

@InProceedings{q-diffusion,
  author={Li, Xiuyu and others},
  title={Q-Diffusion: Quantizing Diffusion Models},
  booktitle={Proceedings of the IEEE/CVF International Conference on Computer Vision (ICCV)},
  month={October},
  year={2023},
  pages={17535-17545}
}

@article{PTQ4DiT,
  title={PTQ4DiT: Post-training Quantization for Diffusion Transformers},
  author={Wu, Junyi and others},
  journal={arXiv preprint arXiv:2405.16005},
  year={2024}
}

@article{qdrop,
title={QDrop: Randomly Dropping Quantization for Extremely Low-bit Post-Training Quantization},
author={Wei, Xiuying and others},
journal={arXiv preprint arXiv:2203.05740},
year={2022}
}

@article{DiTAS,
  title={DiTAS: Quantizing Diffusion Transformers via Enhanced Activation Smoothing},
  author={Dong, Zhenyuan and others},
  journal={arXiv preprint arXiv:2409.07756},
  year={2024}
}

@INPROCEEDINGS{stable_diffusion,
  author={Rombach, Robin and others},
  booktitle={2022 IEEE/CVF Conference on Computer Vision and Pattern Recognition (CVPR)}, 
  title={High-Resolution Image Synthesis with Latent Diffusion Models}, 
  year={2022},
  volume={},
  number={},
  pages={10674-10685},
  doi={10.1109/CVPR52688.2022.01042}}

@article{ViT,
  title={An image is worth 16x16 words: Transformers for image recognition at scale},
  author={Dosovitskiy, Alexey},
  journal={arXiv preprint arXiv:2010.11929},
  year={2020}
}

@article{zeroquant,
  title={Zeroquant: Efficient and affordable post-training quantization for large-scale transformers},
  author={Yao, Zhewei and others},
  journal={Advances in Neural Information Processing Systems},
  volume={35},
  pages={27168--27183},
  year={2022}
}

@inproceedings{ptq4vit,
  title={Ptq4vit: Post-training quantization for vision transformers with twin uniform quantization},
  author={Yuan, Zhihang and others},
  booktitle={European conference on computer vision},
  pages={191--207},
  year={2022},
  organization={Springer}
}

@inproceedings{ptq4dm,
  title={Post-training quantization on diffusion models},
  author={Shang, Yuzhang and others},
  booktitle={Proceedings of the IEEE/CVF conference on computer vision and pattern recognition},
  pages={1972--1981},
  year={2023}
}

@article{ptqd,
  title={Ptqd: Accurate post-training quantization for diffusion models},
  author={He, Yefei and others},
  journal={Advances in Neural Information Processing Systems},
  volume={36},
  year={2024}
}

@inproceedings{smoothquant,
  title={Smoothquant: Accurate and efficient post-training quantization for large language models},
  author={Xiao, Guangxuan and others},
  booktitle={International Conference on Machine Learning},
  pages={38087--38099},
  year={2023},
  organization={PMLR}
}

@article{brecq,
  title={BRECQ: Pushing the Limit of Post-Training Quantization by Block Reconstruction},
  author={Li, Yuhang and others},
  journal={arXiv preprint arXiv:2102.05426},
  year={2021}
}

@inproceedings{li2023repq,
  title={Repq-vit: Scale reparameterization for post-training quantization of vision transformers},
  author={Li, Zhikai and others},
  booktitle={Proceedings of the IEEE/CVF International Conference on Computer Vision},
  pages={17227--17236},
  year={2023}
}

@article{DDPM,
  title={Denoising diffusion probabilistic models},
  author={Ho, Jonathan and others},
  journal={Advances in neural information processing systems},
  volume={33},
  pages={6840--6851},
  year={2020}
}

@ARTICLE{tcsvt_recon,
  author={Chu, Tianshu and others},
  journal={IEEE Transactions on Circuits and Systems for Video Technology}, 
  title={Improving the Post-Training Neural Network Quantization by Prepositive Feature Quantization}, 
  year={2024},
  volume={34},
  number={4},
  pages={3056-3060},
  doi={10.1109/TCSVT.2023.3311923}}

@ARTICLE{tcsvt_dm_speedup1,
  author={Song, Jiaxun and others},
  journal={IEEE Transactions on Circuits and Systems for Video Technology}, 
  title={Torch-Advent-Civilization-Evolution: Accelerating Diffusion Model for Image Restoration}, 
  year={2025},
  volume={35},
  number={2},
  pages={1478-1491},
  doi={10.1109/TCSVT.2024.3470888}}

@ARTICLE{tcsvt_dm_speedup2,
  author={Wang, Chenyu and others},
  journal={IEEE Transactions on Circuits and Systems for Video Technology}, 
  title={Denoising Reuse: Exploiting Inter-frame Motion Consistency for Efficient Video Generation}, 
  year={2025},
  volume={},
  number={},
  pages={1-1},
  doi={10.1109/TCSVT.2025.3548728}}

@ARTICLE{tcsvt_BinaryViT,
  author={Xiao, Junrui and others},
  journal={IEEE Transactions on Circuits and Systems for Video Technology}, 
  title={BinaryViT: Toward Efficient and Accurate Binary Vision Transformers}, 
  year={2025},
  volume={35},
  number={1},
  pages={195-206},
  doi={10.1109/TCSVT.2024.3457610}}

@ARTICLE{tcsvt_dm_speedup3,
  author={Lu, Siqi and others},
  journal={IEEE Transactions on Circuits and Systems for Video Technology}, 
  title={Speed-Up DDPM for Real-Time Underwater Image Enhancement}, 
  year={2024},
  volume={34},
  number={5},
  pages={3576-3588},
  doi={10.1109/TCSVT.2023.3314767}}

@INPROCEEDINGS{efficientdm,
  title={EfficientDM: Efficient Quantization-Aware Fine-Tuning of Low-Bit Diffusion Models},
  author={He, Yefei and others},
  booktitle={International Conference on Learning Representations},
  year={2024}
}

@INPROCEEDINGS{mixedprecision,
  title={Mixed-precision neural network quantization via learned layer-wise importance},
  author={Tang, Chen and others},
  booktitle={European conference on computer vision},
  pages={259--275},
  year={2022},
  organization={Springer}
}

@INPROCEEDINGS{Q-DiT,
  title     = {Q-DiT: Accurate Post-Training Quantization for Diffusion Transformers}, 
  author    = {Lei Chen and others},
  booktitle = {CVPR},
  year      = {2025}
}

\begin{IEEEbiography}[{\includegraphics[width=1in,height=1.5in,clip,keepaspectratio]{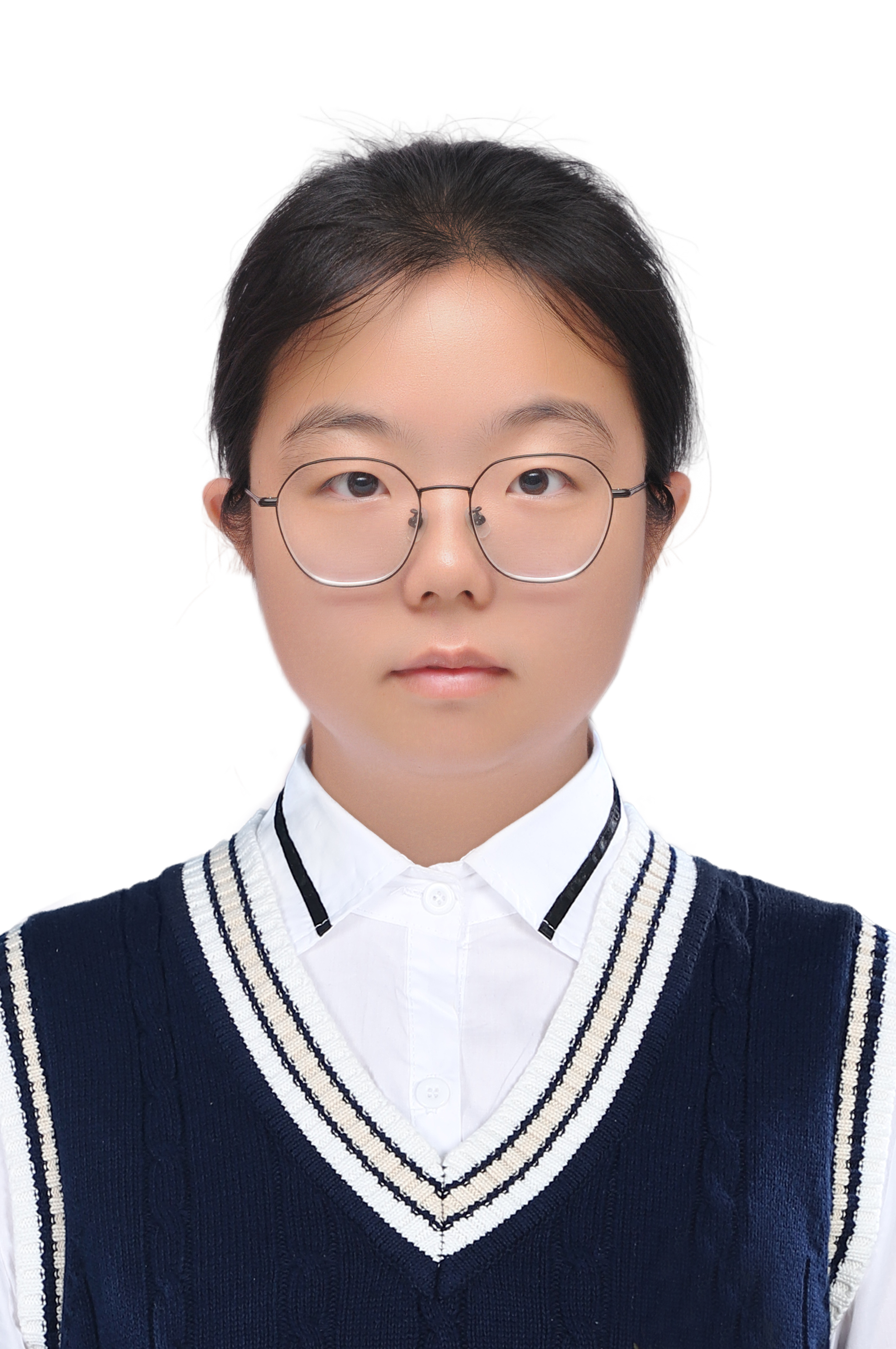}}]{Xinyan Liu} 
received her B.S. degree in microelectronics science and engineering from Nanjing University, Nanjing, China, in 2023. She is currently pursuing the M.S. degree in the School of Electronic Science and Engineering, Nanjing University, China. Her current research interests include model compression and hardware acceleration.
\end{IEEEbiography}

\begin{IEEEbiography}[{\includegraphics[width=1in,height=1.5in,clip,keepaspectratio]{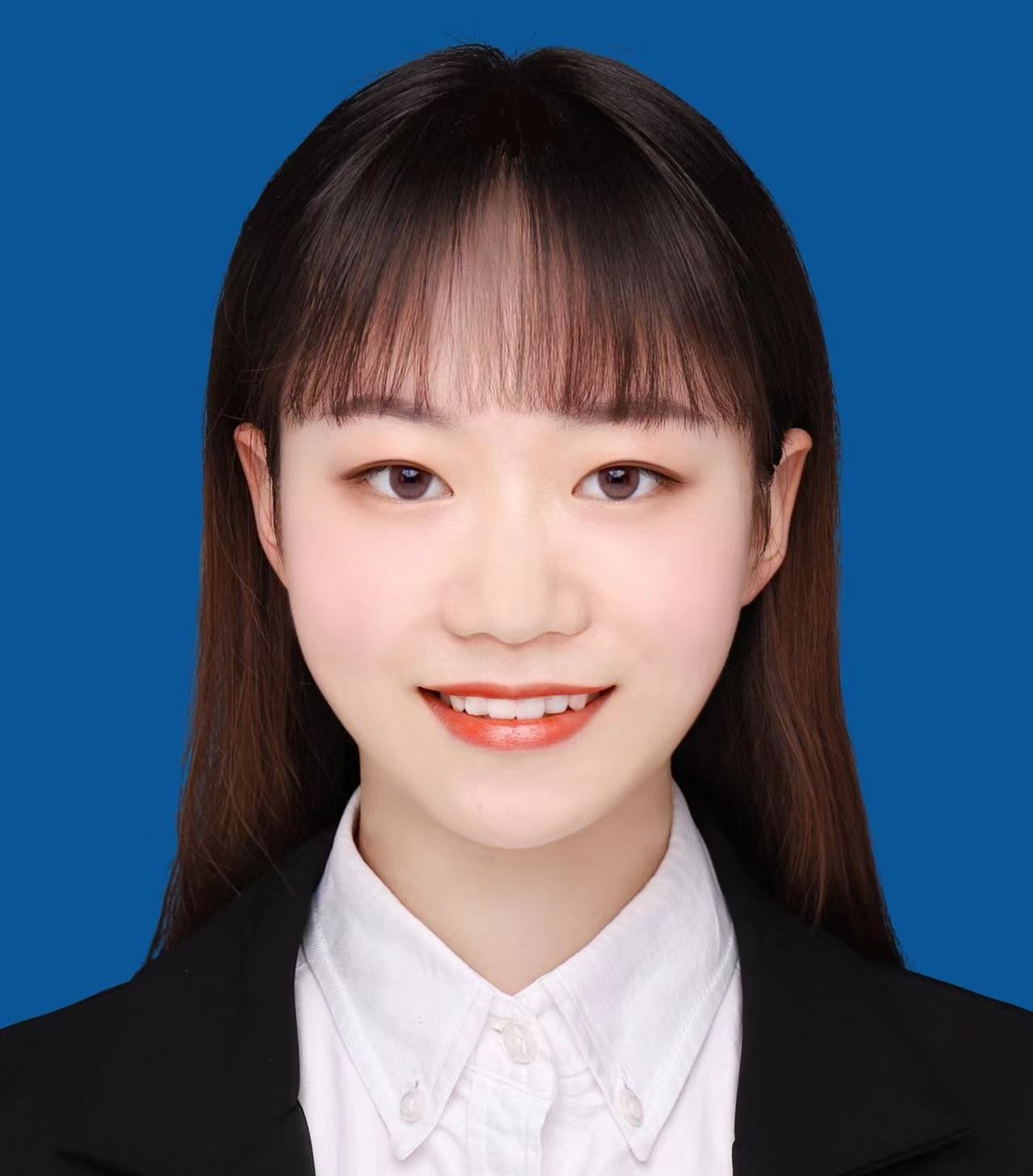}}]{Huihong Shi} 
received her B.S. degree in Communication Engineering from Jilin University in 2020 and her Ph.D. degree in Electronic Science and Engineering from Nanjing University in 2025. Her research interests focus on building efficient and ubiquitous machine learning (ML) systems through algorithm-hardware co-design.
\end{IEEEbiography}

\begin{IEEEbiography}[{\includegraphics[width=1in,height=1.5in,clip,keepaspectratio]{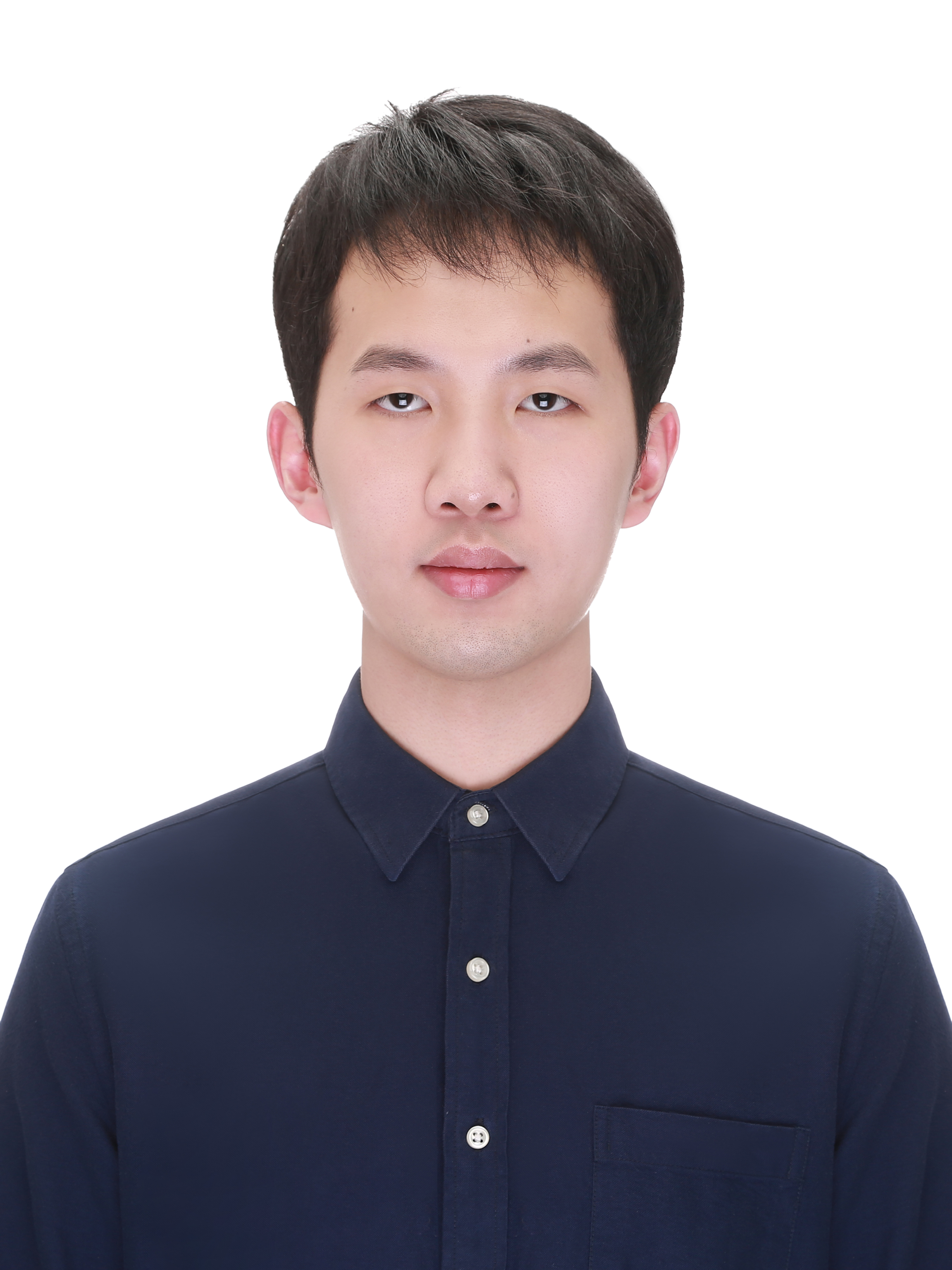}}]{Yang Xu}

received the B.S. degree in electronic engineering from Jilin University, Changchun, China, in 2022 and the M.S. degree from the School of Electronic Science and Engineering, Nanjing University, China, in 2025. His current research interests include neural architecture search and efficient accelerators for deep learning.
\end{IEEEbiography}

\begin{IEEEbiography}[{\includegraphics[width=1in,height=1.25in,clip,keepaspectratio]{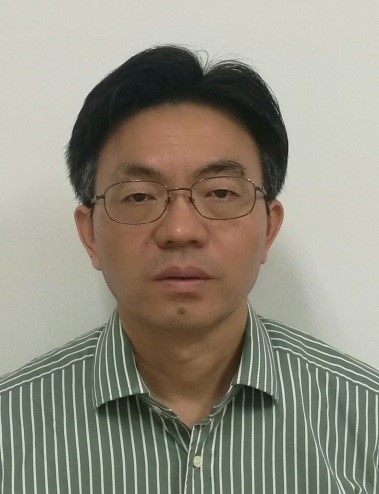}}]{Zhongfeng Wang}
(Fellow, IEEE) received both the B.E. and M.S. degrees in the Dept. of Automation at Tsinghua University, Beijing, China, in 1988 and 1990, respectively. He obtained the Ph.D. degree from the University of Minnesota, Minneapolis, in 2000. He has been working for Nanjing University, China, as a Distinguished Professor since 2016. Previously he worked for Broadcom Corporation, California, from 2007 to 2016 as a leading VLSI architect. Before that, he worked for Oregon State University and National Semiconductor Corporation.
	
Dr. Wang is a world-recognized expert on Low-Power High-Speed VLSI Design for Signal Processing Systems. He has published over 450 technical papers with multiple best paper awards received from the IEEE technical societies, among which is the VLSI Transactions Best Paper Awards of 2007 and 2025. He has edited one book VLSI and held more than 20 U.S. and China patents. In the current record, he has had many papers ranking among top 25 most (annually) downloaded manuscripts in IEEE Trans. on VLSI Systems. In the past, he has served as Associate Editor for IEEE Trans. on TCAS-I, T-CAS-II, and T-VLSI for many terms. He has also served as TPC member and various chairs for tens of international conferences. Moreover, he has contributed significantly to the industrial standards. So far, his technical proposals have been adopted by more than fifteen international networking standards. In 2015, he was elevated to the Fellow of IEEE for contributions to VLSI design and implementation of FEC coding. His current research interests are in the area of Optimized VLSI Design for Digital Communications and Deep Learning.
\end{IEEEbiography}

\end{document}